\documentstyle[aps,twocolumn,floats,prl]{revtex}
\input epsf

\newcommand{\be}{\begin{eqnarray}}      \newcommand{\ee}{\end{eqnarray}}
\newcommand{\rf}[1]{~(\ref{#1})}        \newcommand{\ct}[1]{~\cite{#1}}
\newcommand{\lb}[1]{\label{#1}}         \newcommand{\nn}{\nonumber}

\def\t{\tau}      \def\d{\delta}   \def\dd{\delta}  \def\D{\Delta}
\def\c{c_s}       
\def\phr{\phi_r}  \def\phc{\phi_c}

\begin{document}

\twocolumn[\hsize\textwidth\columnwidth\hsize\csname
@twocolumnfalse\endcsname

\title{Position-Space Description of the Cosmic Microwave Background
           \\ and Its Temperature Correlation Function}

\author{Sergei Bashinsky$^1$ and\, Edmund Bertschinger$^2$}

\address{$^1$Center for Theoretical Physics, $^2$Center for Space Research, and
         $^{1,2}$Department of Physics,\\
         Massachusetts Institute of Technology, Cambridge, Massachusetts 02139}
\maketitle

\begin{abstract}
We suggest that the cosmic microwave background (CMB) temperature
correlation function $C(\theta)$ as a function of angle
provides a direct connection between experimental data and the fundamental 
cosmological quantities.
The evolution of inhomogeneities in the prerecombination universe is studied
using their Green's functions in position space.   We find that a primordial
adiabatic point perturbation propagates as a sharp-edged spherical
acoustic wave.  Density singularities at its wavefronts create a
feature in the CMB correlation function distinguished by a dip at
$\theta \approx 1.2^{\circ}$.  Characteristics of the feature are sensitive
to the values of cosmological parameters, in particular to the total and the
baryon densities.
\end{abstract}
 
\vskip 0.5truecm
]

The cosmic microwave background (CMB) radiation provides the best probe
today of the early universe and a number of fundamental astrophysical
constants\ct{toco,Jaffe,TegZal,Hu00,Dodelson,Bahcall,BET97}.
Dynamical evolution of primordial perturbations manifests itself
in the form of ``acoustic peaks'' in the CMB temperature
angular power spectrum~$C_l$.   After over a decade of studies, the
physical content of the peaks has~become qualitatively
understood\ct{HSS-Nature,HW96,HuSug,WSS-ARev}.
Nevertheless, one has to rely on
standard numerical codes\ct{CMBfast,CAMB} to establish a
quantitative connection between the CMB anisotropy pattern
and cosmological parameters.  This connection is particularly
important now that high-precision CMB measurements have
become reality.

In this letter we show that previously unnoticed interesting
phenomena are unraveled by considering the radiation-matter
dynamics in position rather than Fourier space.
Traditional CMB anisotropy analyses begin with the
Fourier expansion of spatial inhomogeneities and studying
time evolution of individual Fourier modes\ct{Bard80,KS,Mukh}.
The position space approach, while offering a formally equivalent
description, differs in its physical interpretation, associated
calculational methods, and its implications for data analysis. 
As we show here, a new side of CMB
physics is revealed in position space.  One goal of our work
is a deeper physical understanding of both the CMB power spectrum
$C_l$ and the angular correlation function $C(\theta)$.
We demonstrate that acoustic evolution of perturbations
before recombination, as viewed in position space, produces
sharp features in~$C(\theta)$.
These signatures not only provide a simple 
interpretation of the acoustic peaks, but they may also enable 
fast and accurate extraction of cosmological parameters directly 
from~$C(\theta)$.

As a simplified model for the essential physical processes,
we consider the evolution of potential and density perturbations
in gravitationally interacting photon-baryon and cold dark matter
fluids.  We use the conformal Newtonian gauge~\cite{Mukh,MaBert} to
describe gravity and assume adiabatic (isentropic) primordial
fluctuations as preferred by the present data\ct{boom,max}
and the simplest inflationary models.
The photons are assumed to be tightly coupled to electrons
and baryons by Thomson scattering until recombination at redshift
$z_{\rm rec}\sim 1100$.  Neutrinos are treated like photons.  The disregard
of neutrino free-streaming and photon diffusion introduces a $10\%-20\%$
error\ct{Seljak,HSSW} for CMB temperature anisotropy.  
This model is helpful for developing understanding.  Our numerical results
for the correlation function and the angular spectrum are obtained
a full calculation with CMBFAST\ct{CMBfast} including all relevant effects.

Long before recombination during the radiation era,
the growing mode of gravitational potential perturbation
is described in Fourier space by $\phi({\bf k},\t)=
3\left[ \sin (q)/q^3 - \cos (q)/q^2 \right]\phi_0({\bf k})$,
$q\equiv k\c\t$, where $\t$ is conformal time,
${\bf k}$ is the wave vector corresponding to a comoving
coordinate ${\bf r}$, and $\c=1/\sqrt3$ is the speed of sound in
the radiation era\ct{Bard80}.  A delta function primordial
perturbation~$\phi({\bf r},\t\to0)=\dd^{(3)}({\bf r})$
is represented in $k$-space by constant initial amplitude $\phi_0({\bf k})= 1$.
Fourier transforming the resulting $\phi({\bf k},\t)$ we obtain
the three-dimensional Green's function for
the gravitational potential in the radiation dominated epoch,
\be
\phi^{(3)}(r,\t)={3\over4\pi}\,(\c\t)^{-3}\,\theta(\c\t-r)\,,
\lb{radphi3}
\ee
where
$\theta(x)\equiv\{1,\,x> 0;~ 0\ \mbox{otherwise}\}$.
Eq.\rf{radphi3} illustrates the essential property of the position-space
Green's functions---the perturbation is localized within the acoustic
sphere, with a sharp edge at the sound horizon
$r=S(\t)\equiv\int_0^{\t}\c(\t')\,d\t'$.

When cold dark matter (CDM) and baryons are taken into account,
we split the total gravitational potential $\phi=\phr+\phc$
into a part connected to the photon-baryon plasma, $\phr$, and a part
due to the CDM, $\phc$\,.
The sources of the potentials~$\phr$ and $\phc$ are given respectively
by the radiation-baryon and CDM energy densities relative to the
net zero-momentum hypersurfaces, the analogues of Bardeen's\ct{Bard80}
gauge-invariant density variable~$\epsilon_m$.
The potentials  $\phr$ and $\phc$ unambiguously determine
the values of all the density and velocity perturbations
in our model\ct{future}.  Their evolution is described by a pair
of coupled equations,
\be
\ddot\phr+{\textstyle\sum\nolimits_{i=r,c}}(A_{ri}\dot\phi_i+B_{ri}\phi_i)
    &=&\c^2\,\nabla^2\phr\ , \lb{phieqs} \\
\ddot\phc+{\textstyle\sum\nolimits_{i=r,c}}(A_{ci}\dot\phi_i+B_{ci}\phi_i)&=&0
\nn
\ee
where $A_{ij}$ and $B_{ij}$ are simple rational functions of~$\t$\ct{future}.

The first of eqs.\rf{phieqs} shows that a primordial isentropic perturbation
at a point propagates as a spherical acoustic wave in the photon-baryon
plasma with the sound speed $\c^2=d p_{\gamma}/d(\rho_{\gamma}+\rho_{\rm b})
=(1/3)/[1+(3\rho_{\rm b})/(4\rho_{\gamma})]$,
where $\rho_{\gamma}$ and $\rho_{\rm b}$ are the photon and baryon
mean energy densities.
As the sound wave passes by, its gravitational effect perturbs
the CDM causing it to evolve as described by
the second of eqs.\rf{phieqs}.  Since perturbations propagate differently
in the radiation and the CDM components, inhomogeneities do not
remain isentropic for~$\t>0$\,.

We integrate eqs.\rf{phieqs} numerically along their
characteristics starting from the radiation-era solution\rf{radphi3}
at $\t_{\rm init}\simeq 10^{-4}\t_{\rm rec}$, and increasing
the space-time grid spacing with time in proportion to the acoustic
horizon size~$S(\t)$.   It is convenient to solve eqs.\rf{phieqs}
in only one spatial dimension with
the initial conditions $\phi^{(1)}(x)\to\dd(x)$ and vanishing
entropy perturbation as $\t\to 0$.
The resulting one-dimensional, or plane-parallel, Green's function
for the total potential,~$\phi^{(1)}(x,\t)$, is related to 
the conventional transfer function $\phi(k,\t)$ as
$\phi(k,\t)=\int \phi^{(1)}(x,\t)\,e^{-ikx}dx$.
The potential Green's function in three dimensions equals
$\phi^{(3)}(r,\t)=-(2\pi r)^{-1}\partial\phi^{(1)}(r,\tau)
/\partial r$.  A superscript~``$(1)$'' will be used for any quantity
(e.g. density perturbation or temperature anisotropy)
obtained from the isentropic initial condition $\phi(x,\tau\to0)=\dd(x)$,
and such quantities will similarly be referred to as ``Green's functions.''
The Fourier transform of any Green's function gives the corresponding
$k$-space transfer function.

Fig.~1 shows the result for an effective temperature Green's
function $\D^{(1)}_{\rm eff}(x)$
describing the combined intrinsic and gravitational redshift contributions
to CMB temperature anisotropy.
\begin{figure}[tb]
\centerline{\epsffile{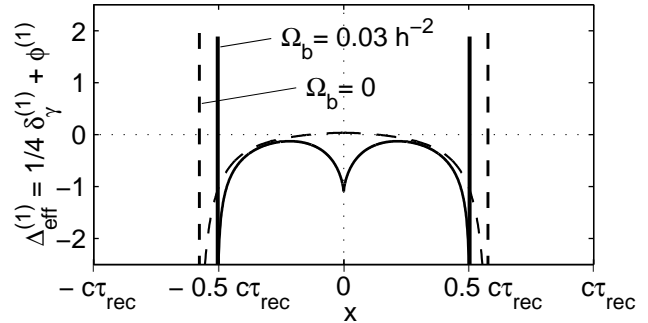}}
\caption{The position space transfer function for the intrinsic
($\frac14\delta_\gamma$) plus gravitational redshift contributions
to the CMB temperature anisotropy $\D T/T$ at recombination, for
$\Omega_{\rm m}=0.35$, $\Omega_{\Lambda}=0.65$, and $h=0.65$.
The vertical spikes represent the wave-front delta function
singularities described in the text.  Baryon drag effects are
evident from comparing the solid and dashed lines.}
\end{figure}
Within our tight coupling assumption, it contains delta function
singularities at the wave fronts given by
$(3/4)\left[1+(3\rho_{\rm b})/(4\rho_\gamma)\right]^{-1/4}
\dd\left(|x|-S\right)$, where $S(\t)\equiv\int_0^{\t}\c(\t')d\t'$.
The contribution of the delta function to
$\int\D^{(1)}_{\rm eff}(x,\t_{\rm rec})\,dx$
exceeds the absolute value of the contribution from the
regular part by $\sim 40\%$,
showing that the singularities play a major
role in CMB anisotropy.

The massive baryons coupled to the radiation fluid reduce the
sound speed that decreases the acoustic horizon at recombination.
They also drag radiation out of the $x=0$ region in Fig.~1
being repelled by a kink in CDM potential~$\phc^{(1)}$ at $x=0$.
This effect causes the prominent ``gully'' in the middle of 
the transfer function in Fig.~1.

The peculiar features of the position space transfer functions
are manifested in the angular correlation function
\be
C(\theta)\equiv  \left\langle\D T({\bf \hat n}_1)
                        \D T({\bf \hat n}_2)\right\rangle=
          \sum_l {(2l+1) \over 4\pi}\, C_lP_l(\cos\theta)\,,
\ee
where $\D T({\bf \hat n}_i)$ is the temperature
anisotropy in the direction ${\bf \hat n}_i$, and $\theta$ is the
angle between ${\bf \hat n}_1$ and ${\bf \hat n}_2$\,.
The connection between $C(\theta)$ and the Green's functions
may be illustrated in the large angle limit when
$\D T({\bf \hat n}_i)/T \approx
\D_{\rm eff}({\bf \hat n}_i,\t_{\rm rec}) \equiv
\frac14\d_{\gamma} + \phi$, $\d_{\gamma}\equiv
\d\rho_{\gamma}/\rho_{\gamma}$, for $\theta$ exceeding
a few degrees.  Writing each $\D T({\bf \hat n}_i)$ as the
convolution of the Green's function in Fig.~1 with a primordial
potential field~$\phi_0$, assuming that recombination occurs
instantaneously, and averaging over the primordial
fluctuations with power spectrum $\langle\phi_0(
{\bf k}_1)\phi_0({\bf k}_2)\rangle\propto
\d^{(3)}({\bf k}_1+{\bf k}_2)k^{n_s-4}$, we get
\be
C(\theta)\propto
\int dx_1\,\D_{\rm eff}^{(1)}(x_1)
\int dx_2\,\D_{\rm eff}^{(1)}(x_2)\,
K(x_1+x_2,\theta)\,,
\lb{Cexpl}
\ee
where $K$ is an analytically known function of $x_1+x_2$,
angle $\theta$, the comoving distance to CMB photosphere~$R$,
and the spectral index~$n_s$\ct{future}.
The exact formula for $C(\theta)$ applicable to subdegree scales
should also include Doppler and integrated Sachs-Wolfe
effects and integration along the line of sight\ct{CMBfast}.
Eq.\rf{Cexpl} neglects these contributions but is sufficient
to give a good qualitative description of the effects considered
below.

From eq.\rf{Cexpl}, the  delta function singularities produce
a singular $C(\theta)$ behavior in the vicinity of
$\theta_s \equiv 2\arcsin(S/R)$, where
$S\equiv S(\t_{\rm rec})\simeq 140$~Mpc is
the acoustic horizon size at recombination.
For a flat universe and the scale-invariant power spectrum~$n_s=1$,
$C(\theta)-C(\theta_s)\sim
-\D s\ln\left|\D s\right|$,
$\D s \equiv \sin(\theta/2)-S/R$.
When the two observed points come into acoustic contact
at the critical angle~$\theta_s$,
$dC(\theta_s)/d\theta=+\infty$
in the instantaneous recombination approximation.
At this angular separation, a spherical sound wave emitted at $\tau=0$
from the point halfway between the observed points on the photosphere
has, by recombination, just reached the two observed points separated
by a distance~$2S$.

Paradoxically, the initial acoustic contact anti-correlates
the temperature fluctuations causing a {\it dip} in the correlation
function just below~$\theta_s$\,.  As evident from Fig.~2\,, the dip
is also present in the full result obtained from a CMBFAST calculation
(wide solid line) although smoothed by the Silk damping\ct{Silk_damp} and
the finite thickness
of the recombination photosphere.  An anomaly in the correlation function
at this angular separation was pointed out earlier in\ct{Starob}
and\ct{Sakh_osc}.  At even smaller angles the correlation
function rises steeply.  The dip occurs when the comoving separation of the
observed points on the photosphere is about $2S\simeq 280$~Mpc,
providing a ``standard ruler'' in the early universe.
The comoving distance to the photosphere $R$ and thus the observed angle
$\theta_s$ are determined by the rate of the subsequent
Hubble expansion and by the cosmic curvature, which both depend
on the total density parameter~$\Omega$.
\begin{figure}[t]
\centerline{\epsffile{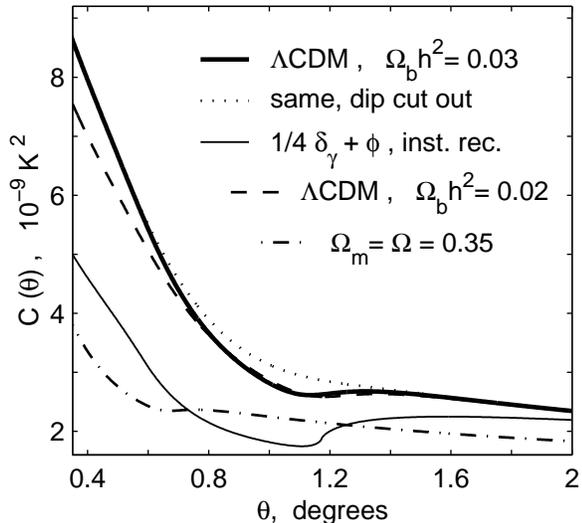}}
\caption{Angular CMB temperature correlation function.
The parameters of the $\Lambda$CDM model are
$\Omega_{\rm m}=0.35$, $\Omega_{\Lambda}=0.65$, $h=0.65$, $n_s=1$.
For this set of parameters and the displayed $\Omega_{\rm b}h^2$ values,
$\theta_s\simeq 1.17^{\circ}~\mbox{and}~1.22^{\circ}$.
The thin solid curve is calculated using eq.~(4).
All the other plots are obtained in full CMBFAST
calculation, except for the dotted line that
artificially cuts the dip for its Fourier analysis in Fig.~4.
The whole feature in~$C(\theta)$ is shifted to smaller
angles in the open model (dash-dotted line).  Baryon density variation
modifies the slope for $\theta<\theta_s/2$
(dashed vs.\ solid line).}
\end{figure}
The dash-dotted line in Fig.~2 demonstrates the shift in the
visible location of the dip for an open model.

What is the physical origin of the dip in the correlation function at
the acoustic contact?   According to our earlier discussion,
$\Delta_{\rm eff}^{(3)}(r)=-(2\pi r)^{-1}\partial\Delta_{\rm eff}^{(1)}
/\partial r$, which contains the derivative of a delta function.
When convolved with the primordial potential field~$\phi_0$,
$\d'(r-S)$ relates $\D T({\bf r})$ to the {\it gradient} of $\phi_0({\bf r})$.
The vectorial nature of~${\bf \nabla} \phi_0$ results in contributions of 
opposite signs to the temperature of two points located at
distance~$S$ in opposite directions.

A larger primordial power spectrum~$n_s$ leads to a greater magnitude
of all small-scale features in the correlation function.
In the instantaneous recombination approximation,
$C(\theta)-C(\theta_s)$ varies as $\D s/(1-n_s)$ for $n_s<1$
and $\D s/|\D s|^{n_s-1}$ for $n_s>1$.
Thus, the dip at $\theta=\theta_s$ is enhanced for $n_s>1$
and suppressed for $n_s<1$.

For $\theta\lesssim\theta_s/2$ the correlation function slope
becomes sensitive to the baryon density~$\Omega_{\rm b}$.
This can be seen from comparing the dashed and the
solid lines in Fig.~2.
At such a small angular separation, the Doppler
and ISW effects provide significant
contribution to the temperature correlation.  Nevertheless, the
$\Omega_{\rm b}$ dependence is largely determined by the same intrinsic
plus gravitational redshift contributions shown in Fig.~1.
It is the interference of the baryon induced ``gully'' in the middle of
one of the transfer functions in eq.\rf{Cexpl}
and the delta function contributions of the other that causes the steeper
$C(\theta)$ rise at $\theta\lesssim\theta_s/2$.
The effect is almost
linearly proportional to~$\Omega_{\rm b}h^2$ and may be useful
for measuring the baryon density value using CMB data.

The finite extent of the position-space Green's functions
results in oscillations in their Fourier transforms, the
$k$-space transfer functions.  These oscillations appear as the
famous acoustic peaks in the CMB angular spectrum~$C_l$\,.
The characteristic period of the oscillations
is $\D l = \pi R/S \simeq 300$ for $\Omega_{\rm m}=0.35$,
$\Omega_{\Lambda}=0.65$, and $h=0.65$.
The wavefront and $x=0$ Green's function
singularities in Fig.~1 do not
map directly to a particular distinct feature in $C_l$ spectrum
because they are completely delocalized by Fourier transformation.

\begin{figure}[b]
\centerline{\epsffile{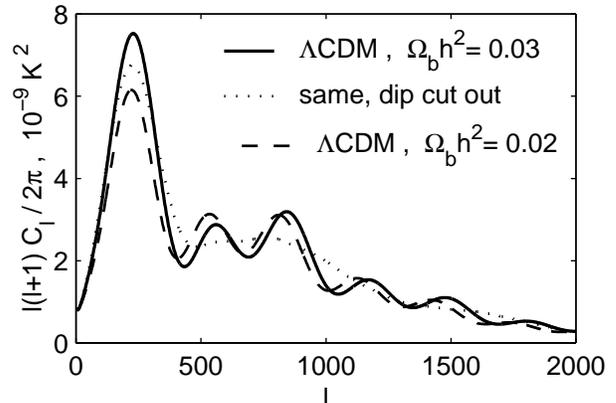}}
\caption{Angular power spectra for the $\Lambda$CDM models of
Fig.~2. The dotted line gives $C_l$ corresponding to the correlation
function of the first model with the dip cut out.}
\end{figure}
An interesting connection between features in $C(\theta)$
and the acoustic peaks in $C_l$ is established by artificially
cutting out the dip in $C(\theta)$ and replacing it  by a smooth
interpolation with no local minimum as shown by the dotted line
in Fig.~2.  Fig.~3. shows that this operation eliminates all the
acoustic peaks besides the first one.
The first peak is closely connected with the sharp rise
in~$C(\theta)$ at $\theta<\theta_s$.  Increasing the baryon density
we increase the magnitude of the slope of~$C(\theta)$.
The magnitude of the first
peak grows accordingly (dashed vs.\ solid lines in Fig.~3) leading
to the known\ct{HuSug} reduction in the ratio of second to first
acoustic peaks.

In conclusion, the Green's function method provides a promising
alternative to the conventional Fourier analysis
of perturbations in the early universe.  It offers
new analytical methods and intuition for the
familiar phenomena such as the acoustic oscillations and
the dependence of peak ratios on $\Omega_{\rm b}h^2$.
It also reveals new features that are localized and easily noticeable in
position space but which were stretched over a multitude of wavenumbers
in the Fourier domain.  In particular, the singularities of the
Green's functions lead to potentially measurable features in the
CMB angular correlation function.

Transfer functions are computed faster in position space than
in Fourier space because the integration effort is concentrated
within compact regions of space interior to the acoustic horizon.
Computation of $\D^{(1)}_{\rm eff}(x,\tau_{\rm rec})$ in Fig.~1
takes less than one second for $0.01\%$ numerical accuracy on
a PC.  We are working to extend the method to integration
of the Boltzmann equation for free-streaming neutrinos and for
photons after recombination.

Experimental data are collected in position space.  As we have
attempted to demonstrate, position space is also more appropriate
for describing the acoustic
dynamics of CMB radiation.  It may thus seem natural to remain in
position space when comparing observational results with theoretical
predictions.
The utility of the correlation function $C(\theta)$ for data analysis
has been recently considered by Szapudi {\it et al.}\ct{SPPSB}.
Data analysis using the correlation function is
complicated  by the covariance of observational
estimates of $C(\theta)$\ct{selbert}.  Such correlations are eliminated
for estimates of $C_l$ made with uniformly-sampled all-sky maps.
Nonetheless, the correlation function may still be a valuable
experimental tool because it concentrates much of the cosmologically
essential information into localized features.  This may enable
data processing efforts, and observing strategy, to be focused
on the most relevant angular features.

We thank J.~Bond, K.~Burgess, D.~Pogosyan, and A.~Shirokov for helpful
discussion and gratefully acknowledge the hospitality of
CITA where part of this work was performed.  We thank an 
anonymous referee for the suggestion to consider~$n_s$ variation.  
Support was provided by NSF grant AST-9803137 and by the 
U.~S.\ Department of Energy under cooperative research agreement 
DF-FC02-94ER40818.

\end{document}